\documentclass[12pt]{article}

\usepackage{amsfonts,amssymb,amsmath}
\usepackage{graphicx}
\usepackage{url}
\usepackage{natbib}

\newcommand{\Qx}{ \mathbb{Q} }
\newcommand{\E}[1]{\mathbb{E}\!\left[\,#1\,\right]}
\newcommand{\Ex}[2]{\mathbb{E}_{#1}\!\left[\,#2\,\right]}
\newcommand{\eqdef}{\mathrel{\mathop:}=}
\newcommand{\cva}{\mbox{CVA}}
\newcommand{\rec}{\mbox{R{\tiny EC}}}
\newcommand{\cash}{\Pi}
\newcommand{\lgd}{\mbox{L{\tiny GD}}}
\newcommand{\ind}[1]{1_{\{#1\}}}

\newcommand{\brcva}{\mathrm{BCVA}}
\newcommand{\brccva}{\mathrm{BCCVA}}
\newcommand{\ccva}{\mathrm{CCVA}}
\newcommand{\cdva}{\mathrm{CDVA}}

\newtheorem{theorem}{Theorem}[section]
\newtheorem{remark}[theorem]{Remark}

\begin{document}

\title{
{\bf\Large\shortstack{Collateral Margining in Arbitrage-Free\medskip\\Counterparty Valuation Adjustment\medskip\\including Re-Hypotecation and Netting}}
}

\author{
Damiano Brigo\thanks{Dept. of Mathematics, King's College, London, E-mail: {\tt damiano.brigo@kcl.ac.uk}} , \
Agostino Capponi\thanks{Dept. of Industrial Engineering, Purdue University, E-mail:{\tt capponi@purdue.edu}}, \\
Andrea Pallavicini\thanks{Banca Leonardo, Milano, Italy, {\tt andrea.pallavicini@bancaleonardo.com}}, \
Vasileios Papatheodorou\thanks{Independent Consultant, London, UK, E-mail: {\tt vpapatheo@gmail.com }.
}
}

\date{\small First Version: Sep 3, 2010. This Version: \today}

\maketitle

\vspace{-1cm}

\begin{abstract}
This paper generalizes the framework for arbitrage-free valuation of bilateral counterparty risk to the case where collateral is included, with possible re-hypotecation. We analyze how the payout of claims is modified when collateral margining is included in agreement with current ISDA documentation. We then specialize our analysis to interest-rate swaps as underlying portfolio, and allow for mutual dependences between the default times of the investor and the counterparty and the underlying portfolio risk factors. We use arbitrage-free stochastic dynamical models, including also the effect of interest rate and credit spread volatilities. The impact of re-hypotecation, of collateral margining frequency and of dependencies on the bilateral counterparty risk adjustment is illustrated with a numerical example.
\end{abstract}

{\bf JEL classification code: G13. \\ \indent AMS classification codes: 60J75, 91B70}

\medskip

{\bf Keywords:} Counterparty Risk, Bilateral CVA, Collateral Management, Collateral Re-hypothecation, Close-Out Amount, Margining Procedure, Netting Rules, Hybrid Products, Correlation, Risk Neutral Valuation, Default Risk, Interest Rate Models, Default Intensity Models.{\let\thefootnote\relax\footnotetext{\par{\noindent The opinions expressed in this work are solely those of the four authors and do not represent in any way those of their current and past employers.}}}

\newpage
{\small \tableofcontents}
\newpage

\pagestyle{myheadings}
\markboth{}{{\footnotesize  Brigo, Capponi, Pallavicini, Papatheodorou: Collateral modeling for CVA}}

\section{Introduction}

\subsubsection*{Unilateral CVA}

Basel II defines the counterparty credit risk as the risk that the counterparty to a transaction could default before the final settlement of the transaction. If the party who defaulted is a debtor to the other party in the transaction at the default time, then this would result in an economic loss for the non-defaulted party.

Situations where only default of one of the two parties is taken into account are referred to as unilateral counterparty risk. In such cases only the default of one name impacts valuation. The resulting adjustment to the otherwise default-free price of the deal, computed by the party whose default is not considered, is termed unilateral Credit Valuation Adjustment (UCVA). Unilateral CVA has been considered for example in \cite{sorensen} and in \cite{BieleckiRutkbook}, among others. Pricing of UCVA under netting is considered for example in \cite{BrigoMas}, whereas UCVA with collateral is discussed in some stylized cases and for basic products such as forward contracts in \cite{Cher05}. Precise pricing of UCVA on several asset classes with full arbitrage free dynamic models and wrong way risk is then considered in \cite{BrigoPall07} (Interest rate swaps under netting and derivatives), \cite{BrigoChourBakkar} (Oil swaps), and \cite{Brigo08} (Credit, and CDS in particular), although these works do not account for collateralization.

Despite the unilateral CVA being considered in the beginning by several institutions, Basel II had recognized the bilateral nature of counterparty risk, mentioning that unlike a firm's exposure to credit risk through a loan, where the exposure to credit risk is unilateral and only the lending bank faces the risk of loss, the counterparty credit risk creates a bilateral risk of loss. The market value of the transaction can be positive or negative to either counterparty to the transaction. It follows that if both parties may default, then the counterparty risk calculation becomes a bilateral one.

Indeed, the ongoing financial crisis has made quite clear that the unilateral assumption is not realistic. In particular, this assumption has been put into question by the seven credit events on financials that happened in one month during the period going from September, 7 2008 to October, 8 2008, namely the credit events on Fannie Mae, Freddie Mac, Lehman Brothers, Washington Mutual, Landsbanki, Glitnir and Kaupthing.

\subsubsection*{Bilateral CVA}

The bilateral counterparty risk feature was first considered in the literature by \cite{DuffieHuang}, who present a model for valuing claims subject to default by both contracting parties, such as swap and forward contracts. In their approach, when counterparties have different default risk, the promised cash flows of the swap are discounted using a switching discount rate that, at any given state and time, is equal to the discount rate of the counterparty for whom the swap is currently out of the money. A general formula for bilateral counterparty risk evaluation that carefully takes into account the appropriated cash flows in every event was given in \cite{BieleckiRutkbook}, where the analysis then focuses on the particular case of interest rate swap contracts. The importance of considering the bilateral nature is again mentioned in the excellent exposition of \cite{CannDuff} on the mechanics and valuation of counterparty risk, where it is stated that both parties may face exposures depending on the value of the positions they hold against each other. \cite{Picoult} also reports a detailed formula with discussion of bilateral default risk. However, this is a simplified formula that does not check which counterparty defaults first but simply subtracts the two unilateral CVA adjustments seen by the two parties without considering the actual sequence of default times. This involves some double counting. \cite{BrigoCap10} develop a fully rigorous formula, showing that the bilateral counterparty risk adjustment (BCVA) computed by one of the two parties is obtained as a difference of two terms: a Credit Valuation Adjustment driven by the default event of the other party, and a Debit Valuation Adjustment (DVA) driven by the default of the party that is doing the calculation. They include the proper sequence of default events into the formula to avoid double counting. See also \cite{Gregory}. The DVA term involves some counterintuitive features, such as the fact that when the credit quality of a party worsens its mark to market increases.

The ongoing financial crisis has led the Basel Committee to revisit the guidelines to follow for OTC derivatives transactions, moving towards a new set of rules commonly called ``Basel III", and reviewed in \cite{basel3}. Beside stressing the need to capture correctly the dependence between market and credit risks, also known as wrong- and right-way risk, which was not adequately incorporated into the Basel II framework, they proposed several other amendments. Those include extending the margin period of risk for OTC derivatives, increasing the incentives to use central counterparties to clear trades, and enhancing the controls regarding the re-hypothecation and re-investment of collaterals. \cite{DuffieZhu} address the mitigation of counterparty risk exposure through the use of central clearing counterparties. They show that, adding a central clearing counterparty for credit default swaps can reduce the netting efficiency and lead to an increase in average exposures to counterparty
default.

Basel III also considers CVA as a key element in the analysis of risk. Since during the crisis about two thirds of losses have been due to CVA mark to market and only about one third to actual defaults (see for example \cite{Nathanael}), Basel III is encouraging institutions to include CVA mark to market future simulations in Value at Risk type measures. However, Basel III also suggests a bond equivalent approach to compute CVA in a simple way. This bond equivalent approach is rather unrealistic, focuses on unilateral CVA and again cannot account properly for wrong way risk except through multipliers.

\subsubsection*{Bilateral CVA with Collateral}

In this paper, we study how counterparty risk exposure can be reduced through the use of collateralization. The idea of collateralization of counterparty risk is very similar to the way collateral is used to mitigate lending risk, with collateral used to reduce credit exposure. However, because of the uncertainty of counterparty credit exposure and the bilateral nature of counterparty credit risk, collateral management is much more complex in the case of counterparty risk. Exposure of one counterparty to another changes every day, and to keep the current exposure under control, it is necessary to post collateral frequently, ideally on a daily basis. The
collateral should be used to hedge the exposure that one party has to the other on the default event. The collateral can be in the form of risk-free cash flow or of a (defaultable) asset. In the latter case, it should not be correlated to the value of the transaction, and be liquid, i.e. sold quicky and easily if the need arises.

We develop an arbitrage-free valuation framework for bilateral counterparty risk adjustments, inclusive of collateralization.  We provide model independent formulas that give the bilateral collateralized credit valuation adjustment (abbreviated throughout the paper with BCCVA) for portfolios exchanged between a default risky investor and a default risky counterparty. Such formulas are given by the sum of option payoff terms, where each term depends on the netted exposure, i.e. the difference between the on-default exposure and the on-default collateral account. We consider both the case when collateral is a risk-free asset kept into a segregate account and only used upon default occurrence to net exposure, and also the case when collateral can be lent or re-hypothecated before default occurrence, thus making the party who posted collateral an unsecured creditor.

For the moment, we leave aside some issues linked to counterparty risk evaluation, which may be relevant in particular settings; among them we cite  funding costs, collateral dispute resolutions, and independent amounts. Since these problems are currently under active investigation by ISDA to tune the Master Agreement and Credit Support Annexes in a post-crisis scenario, we prefer to address them in a further development of our work waiting for ISDA recommendation on definitions and relative market practice. We also leave aside the inclusion of features such as goodwill, for which we refer to \cite{Kenyon}.

This paper generalizes the framework for risk-neutral valuation of bilateral counterparty risk introduced in \cite{BrigoCap10}, who do not model the impact of collateralization. We then specialize our analysis to interest-rate payouts as underlying portfolio, and allow for correlation between the default times of the investor, counterparty and underlying portfolio risk factors. By following \cite{BrigoPallPap} we use arbitrage-free stochastic dynamical models and consider the following dependencies:

\begin{itemize}
\item Dependence between default of the counterparty and default of the investor;
\item Correlation between the underlying (interest rates) and the counterparty credit spread;
\item Correlation between the underlying (interest rates) and the investor credit spread;
\end{itemize}

The rest of the paper is organized as follows. Section \ref{sec:coll form} introduces the collateral account process, and develops a formula for computing the bilateral collateralized credit value adjustment (abbreviated throughout the paper with BCCVA), i.e. including counterparty risk and collateralization. Section \ref{sec:examples} presents some examples of collateralization mechanisms. Section \ref{sec:IRS} presents an application of the bilateral CVA formula to interest rates swap contracts, illustrating the impact of the collateral frequency on the bilateral CVA adjustment. Section \ref{sec:Conclusions} concludes the paper.

\section{Collateralized Credit Value Adjustment}
\label{sec:coll form}

Risk neutral evaluation of counterparty risk in presence of collateral management can be difficult, due to the complexity of clauses. There are only a few papers in the literature dealing with it, among them \cite{Cher05}, \cite{Assefa09}, and \cite{Alavian08}. \cite{Assefa09} consider a highly stylized model for the collateral process without accounting for minimum transfer amounts, collateral thresolds, and assume that the collateral account is risk-free and cannot be re-hypothecated. \cite{Alavian08} discuss features such as minimum transfer amount and collateral thresholds and give model independent formulas for the counterparty exposure, netted of collateralization, but again assuming the collateral is a risk-free asset.

The objective of this section is to provide a model independent formula for the counterparty value adjustment, inclusive of collateralization mitigation, and allowing for default risk of both parties. In the next subsections, we develop a general framework for computing such quantity, and along the way we show the relation between the mathematical formulation and the \cite{ISDA1}.

The rest of the section is organized as follows. Subsection \ref{sec:mathsetup} introduces the mathematical setup. Subsection \ref{sec:closeout} discusses the close-out netting rules and introduces the concept of counterparty and investor on-default exposure. Subsection \ref{sec:collrehyp} discusses how the collateral account can be used throughout the life of the transaction, for example it can be re-hypothecated. Subsection \ref{subs:formulasCVA} provides model-independent formulas which follow naturally from the written standards and contractual rules described in the earlier sections. Subsection \ref{subs:closeoutEval} discusses the issues of calculating on-default exposures.

\subsection{Mathematical Setup}
\label{sec:mathsetup}

We refer to the two names involved in the financial contract and subject to default risk as
\begin{eqnarray}
\nonumber \textrm{investor } & \rightarrow & \textrm{name ``I''} \\
\nonumber \textrm{counterparty } & \rightarrow & \textrm{name ``C''}
\label{eq:three_names}
\end{eqnarray}

We denote by $\tau_I$ and $\tau_C$ respectively the default times of the investor and counterparty. We place ourselves in a probability space $(\Omega,\mathcal{G},\mathcal{G}_t,\mathbb{Q})$. The filtration $\mathcal{G}_t$ models the flow of information of the whole market, including defaults and $\Qx$ is the risk neutral measure. This space is endowed also with a right-continuous and complete sub-filtration $\mathcal{F}_t$ representing all the observable market quantities but the default event, thus $\mathcal{F}_t\subseteq\mathcal{G}_t:=\mathcal{F}_t\vee\mathcal{H}_t$. Here, $\mathcal{H}_t=\sigma(\{\tau_I\leq u \} \vee \{\tau_C\leq u\} :u\leq t)$ is the right-continuous filtration generated by the default events,
either of the investor or of his counterparty (and of the reference credits if the underlying portfolio is credit sensitive).

Let us call $T$ the final maturity of the payoff which we need to evaluate and define the stopping time
\begin{equation}
\label{eq:stopping_time}
\tau = \tau_I \wedge \tau_C
\end{equation}

We define the \textit{collateral account} $C_t$ to be a stochastic process adapted to the filtration $\mathcal{G}_t$. Intuitively, this means that the collateral account at time $t$ ``knows" the values of
all the market observables up to time $t$, including which entities have defaulted by $t$.

We assume that the collateral account is held by the \textit{collateral taker}, with both investor and counterparty posting or withdrawing collateral during the life of a deal, to or from the collateral account. The other party is the \textit{collateral provider}.  We see all payoffs from the point of view of the investor. Therefore, when $C_t > 0$, this means that by time $t$ the overall collateral account is in favor of the investor and the net posting has been done by the counterparty, meaning that what is in the account at $t$ is the excess of posting done by the counterparty with respect to the investor posting. In this case the collateral account $C_t > 0$ can be used by the investor to reduce his on-default exposure. On the contrary, when $C_t < 0$, this means that the overall collateral account by time $t$ is in favor of the counterparty, and has been net-posted by the investor. In this case collateral can be used by the counterparty to reduce his on-default exposure.

Thus when $C_t > 0$ this means that, at time $t$, the collateral taker is the investor and the collateral provider is the counterparty, whereas in the other case the collateral taker is the counterparty and the collateral provider is the investor.

We assume the collateral account to be a risk-free cash account, although in general it can be any other (defaultable) asset, which can be liquidated at the default time. Further, we assume that the collateral account is opened anew for each new deal and it is closed upon a default event or when maturity is reached. If the account is closed, then any collateral held by the collateral taker would be required to be returned to the originating party. We assume $C_t = 0$ for all $t \leq 0$, and $C_t = 0$, if $t \geq T$.

We call $\cash(u,s)$ the net cash flows of the claim under consideration  (not including the collateral account) without investor or counterparty default risk between time $u$ and time $s$, discounted back at $u$, as seen from the point of view of the investor. We denote by $\Pi^D(u,s,C)$ the analogous net cash flows of the claim under counterparty and investor default risk, and inclusive of collateral netting.
%Therefore, when $\cash(u,s) > 0$ or $\Pi^D(u,s,C) > 0$, the investor is creditor to the counterparty, while the opposite is true when  $\cash(u,s) < 0$ or $\Pi^D(u,s,C) < 0$.
The \emph{counterparty valuation adjustment in presence of collateralization} is given by
\[
\brccva(t,T,C) := \Ex{t}{{\Pi^D}(t,T, C)} - \Ex{t}{\Pi(t,T)}
\]

In order to evaluate the CVA inclusive of collateralization, we need to express $\Pi^D(t,T,C)$ in terms of risk-free quantities, default indicators and collateral. In particular we should describe which operations the investor and the counterparty perform to monitor and mitigate counterparty credit risk, and
which operations, on a default event, the surviving party performs to recover from potential losses.

\begin{remark}{\bf(Collateral Delay and Dispute Resolutions)}

In practice there is a delay between the time when collateral is requested and the time when it gets posted. This is due to collateral settlement rules or to one party (or both parties) disputing on portfolio or collateral pricing.

Typically, the delay is limited to one day, but it may be longer. According to the ISDA Collateral Dispute Resolution Protocol (2009) the parties may agree either on a standard timing schedule (disputes end within three days), or on an extended one (disputes end within nine days). Exceptionally, further delay may take place due to a mutual consent of both parties or due to specific market concerns (total delay cannot exceed thirty days). We do not consider collateral posting delay in this paper leaving this issue for future research.

\end{remark}

\subsection{Close Out Netting Rules}\label{sec:closeout}

The ISDA Market Review of OTC Derivative Bilateral Collateralization Practices (2010) on section 2.1.1. states the following:
\begin{quotation}{\it
The effect of close-out netting is to provide for a single net payment requirement in respect of all the transactions that are being terminated, rather than multiple payments between the parties. Under the applicable accounting rules and capital requirements of many jurisdictions, the availability of close-out netting allows parties to an ISDA Master Agreement to account for transactions thereunder on a net basis
}\end{quotation}

This means that, upon the occurrence of a default event, the parties should terminate all transactions and do a netting of due cash-flows. Moreover, the ISDA Credit Support Annex, subject to New York Law, on paragraph 8 states
\begin{quotation}{\it
The Secured Party will transfer to the Pledgor any proceeds and posted credit support remaining after liquidation, and/or set-off after satisfaction in full of all amounts payable by the Pledgor with respect to any obligations; the Pledgor in all events will remain liable for any amounts remaining unpaid after any liquidation and/or set-off.
}\end{quotation}

This means that the surviving party should evaluate the transactions just terminated, due to the default event occurrence, and claim for a reimbursement only after the application of netting rules, inclusive of collateral accounts. We can find similar clauses also in CSAs subject to different laws.

The ISDA Master Agreement defines the term \emph{close-out amount} to be the amount of the losses or costs of the surviving party would incur in
replacing or in providing for an economic equivalent at the time when the counterparty defaults. Notice that the close-out amount is not a symmetric quantity w.r.t. the exchange of the role of two parties, since it is valued by one party after the default of the other one.

The replacing counterparty may ask the surviving party to post more than the exposure to the old defaulted counterparty to compensate for liquidity, or the deteriorated credit quality of the surviving party. Instead of the close-out amount we introduce the \emph{on-default exposure}, namely the price of the replacing transaction or of its economic equivalent. We distinguish between on-default exposure of investor to counterparty and of counterparty to investor at time $t$, and denote it as follows
\begin{itemize}
\item $\varepsilon_{I,t}$ denotes the on-default exposure of the investor to the counterparty at time $t$. A positive value for $\varepsilon_{I,t}$ means that the investor is a creditor of the counterparty.
\item $\varepsilon_{C,t}$ denotes the on-default exposure of the counterparty to the investor at time $t$. A negative value for $\varepsilon_{C,t}$ means that the counterparty is a creditor to the investor.
\end{itemize}

\subsection{Collateral Re-Hypothecation}
\label{sec:collrehyp}

In case of no-default happening, at final maturity the collateral provider expects to get back from the collateral taker the remaining collateral. Similarly, in case of default happening earlier (and assuming the collateral taker before default to be the surviving party), after netting the collateral with the cash flows of the transaction, the collateral provider expects to get back the remaining collateral on the account. However, it is often considered to be important, commercially, for the collateral taker to have relatively unrestricted use of the collateral until it must be returned to the collateral provider. This unrestricted use includes the ability to sell collateral to a third party in the market, free and clear of any interest of the collateral provider. Other uses would include lending the collateral or selling it under a ``repo'' agreement or {\emph{re-hypothecating}} it. Although under the English Deed the taker is not permitted to re-hypothecate the collateral, the taker is allowed to do so under the New York Annex, the English Annex or the Japanese Annex. When the collateral taker re-hypothecates the collateral, then he leaves the collateral provider as an unsecured creditor with respect to collateral reimbursement.

In case of re-hypothecation, the collateral provider must therefore consider the possibility to recover only a fraction of his collateral. If the investor is the collateral taker, we denote the recovery fraction on collateral re-hypothecated by the defaulted investor by $\rec'_I$, while if the counterparty is the collateral taker, then we denote the recovery fraction on collateral re-hypothecated by the counterparty by $\rec'_C$. Accordingly, we define the collateral loss incurred by the counterparty upon investor default by $\lgd'_I = 1 - \rec'_I$ and the collateral loss incurred by the investor upon counterparty default by $\lgd'_C = 1 - \rec'_C$. Typically, the surviving party has precedence on other creditors to get back his collateral, thus $\rec_I \leq \rec_I' \leq 1$, and $\rec_C \leq \rec_C' \leq 1$. Here, $\rec_I$ ($\rec_C$) denote the recovery fraction of the market value of the transaction that the counterparty (investor) gets when the investor (counterparty) defaults.

Notice that the case when collateral cannot be re-hypothecated and has to be kept into a segregate account is obtained by setting $\rec_I' = \rec_C' = 1$.

We need to mention that collateral re-hypothecation has been heavily criticized and is currently debated. See for example the Senior Supervisors Group (2009) report, that observes the following:

\begin{quotation}{\it
Custody of assets and re-hypothecation practices were dominant drivers of contagion, transmitting liquidity risks to other firms. In the United Kingdom, there was no provision of central bank liquidity to the main broker-dealer entity, Lehman Brothers International (Europe), and no agreement was struck to transfer client business to a third-party purchaser. As a result, LBIE filed for bankruptcy while holding significant custody assets that would not be returned to clients for a long time, and therefore could not be traded or easily hedged by clients. In addition, the failure of LBIE exposed the significant risks run by hedge funds in allowing their prime broker to exercise re-hypothecation rights over their securities. Under U.K. law, clients stand as general creditor for the return of such assets.

The loss of re-hypothecated assets and the ``freezing" of custody assets created alarm in the hedge fund community and led to an outflow of positions from similar accounts at other firms. Some firms' use of liquidity from re-hypothecated assets to finance proprietary positions also exacerbated funding stresses.}
\end{quotation}

\subsection{Bilateral CVA Formula under Collateralization}
\label{subs:formulasCVA}

We start by listing all the situations that may arise on counterparty default and investor default events. Our goal is to calculate the present value of all cash flows involved by the contract by taking into account (i) collateral margining operations, and (ii) close-out netting rules in case of default. Notice that we can safely aggregate the cash flows of the contract with the ones of the collateral account, since on contract termination all the posted collateral is returned to the originating party.

\subsubsection*{Collecting CVA Contributions}

We start considering all possible situations which may arise at the default time of the counterparty, which is assumed to default before the investor. In our notation

\[ X^+ = \max(X,0), \ \ X^- = \min(X,0) . \]

 We have

\begin{enumerate}

\item The investor measures a positive (on-default) exposure on counterparty default ($\varepsilon_{I,\tau_C}>0$), and some collateral posted by the counterparty is available ($C_{\tau_C}>0$). Then, the investor exposure is reduced by netting, and the remaining collateral (if any) is returned to the counterparty. If the collateral is not enough, the investor suffers a loss for the remaining exposure. Thus, we have
$$
\ind{\tau=\tau_C<T} \ind{\varepsilon_{I,\tau}>0} \ind{C_\tau>0} (\rec_C(\varepsilon_{I,\tau} - C_\tau)^+ + (\varepsilon_{I,\tau}-C_\tau)^-)
$$

\item The investor measures a positive (on-default) exposure on counterparty default ($\varepsilon_{I,\tau_C}>0$), and some collateral posted by the investor is available ($C_{\tau_C}<0$). Then, the investor suffers a loss for the whole exposure. All the collateral (if any) is returned to the investor if it is not re-hypothecated, otherwise only a recovery fraction of it is returned. Thus, we have
$$
\ind{\tau=\tau_C<T} \ind{\varepsilon_{I,\tau}>0} \ind{C_\tau<0} (\rec_C \varepsilon_{I,\tau} - \rec'_C C_\tau)
$$

\item The investor measures a negative (on-default) exposure on counterparty default ($\varepsilon_{I,\tau_C}<0$), and some collateral posted by the counterparty is available ($C_{\tau_C}>0$). Then, the exposure is paid to the counterparty, and the counterparty gets back its collateral in full.
$$
\ind{\tau=\tau_C<T} \ind{\varepsilon_{I,\tau}<0} \ind{C_\tau>0} (\varepsilon_{I,\tau} - C_\tau)
$$

\item The investor measures a negative (on-default) exposure on counterparty default ($\varepsilon_{I,\tau_C}<0$), and some collateral posted by the investor is available ($C_{\tau_C}<0$). Then, the exposure is reduced by netting and paid to the counterparty. The investor gets back its remaining collateral (if any) in full if it is not re-hypothecated, otherwise he only gets the recovery fraction of the part of collateral exceeding the exposure.
$$
\ind{\tau=\tau_C<T} \ind{\varepsilon_{I,\tau}<0} \ind{C_\tau<0} ( (\varepsilon_{I,\tau} - C_\tau)^- + \rec'_C (\varepsilon_{I,\tau} - C_\tau)^+ )
$$

\end{enumerate}

Symmetrically, we consider all possible situations which can arise at the default time of the investor, which is the earliest to default. We have

\begin{enumerate}

\item The counterparty measures a positive (on-default) exposure on investor default ($\varepsilon_{C,\tau_I}<0$), and some collateral posted by the investor is available ($C_{\tau_I}< 0$). Then, the counterparty exposure is reduced by netting, and the remaining collateral (if any) is returned to the investor. If the collateral is not enough, the investor suffers a loss for the remaining exposure. Thus, we have
$$
\ind{\tau=\tau_I<T} \ind{\varepsilon_{C,\tau}<0} \ind{C_\tau<0} (\rec_I(\varepsilon_{C,\tau} - C_\tau)^- + (\varepsilon_{C,\tau}-C_\tau)^+)
$$

\item The counterparty measures a positive (on-default) exposure on investor default ($\varepsilon_{C,\tau_I}< 0$), and some collateral posted by the counterparty is available ($C_{\tau_I} > 0$). Then, the counterparty suffers a loss for the whole exposure. All the collateral (if any) is returned to the counterparty if it is not re-hypothecated, otherwise only a recovery fraction of it is returned. Thus, we have
$$
\ind{\tau=\tau_I<T} \ind{\varepsilon_{C,\tau}<0} \ind{C_\tau>0} (\rec_I \varepsilon_{C,\tau} - \rec'_I C_\tau)
$$

\item The counterparty measures a negative (on-default) exposure on investor default ($\varepsilon_{C,\tau_I}>0$), and some collateral posted by the investor is available ($C_{\tau_I}<0$). Then, the exposure is paid to the investor, and the investor gets back its collateral in full.
$$
\ind{\tau=\tau_I<T} \ind{\varepsilon_{C,\tau}>0} \ind{C_\tau<0} (\varepsilon_{C,\tau} - C_\tau)
$$

\item The counterparty measures a negative (on-default) exposure on investor default ($\varepsilon_{C,\tau_I}>0$), and some collateral posted by the counterparty is available ($C_{\tau_I}>0$). Then, the exposure is reduced by netting and paid to the investor. The counterparty gets back its remaining collateral (if any) in full if it is not re-hypothecated, otherwise he only gets the recovery fraction of the part of collateral exceeding the exposure.
$$
\ind{\tau=\tau_I<T} \ind{\varepsilon_{C,\tau}>0} \ind{C_\tau>0} ( (\varepsilon_{C,\tau} - C_\tau)^+ + \rec'_I (\varepsilon_{C,\tau} - C_\tau)^- )
$$

\end{enumerate}

Now, we can aggregate all these cash flows, along with cash flows coming from the default of the investor and the ones due in case of non-default, inclusive of the cash-flows of the collateral account. Let $D(t,T)$ denote the risk-free discount factor. By summing all contributions, we obtain

%{\small
\[
\begin{split}
& {\Pi^D}(t,T;C) = \medskip\\
& \quad \quad \,\,\! \ind{\tau>T} \Pi(t,T) \medskip\\
& \quad + \ind{\tau<T} (\Pi(t,\tau) + D(t,\tau)C_\tau) \medskip\\
& \quad + \ind{\tau=\tau_C<T} D(t,\tau) \ind{\varepsilon_{I,\tau}<0} \ind{C_\tau>0} (\varepsilon_{I,\tau} - C_\tau) \medskip\\
& \quad + \ind{\tau=\tau_C<T} D(t,\tau) \ind{\varepsilon_{I,\tau}<0} \ind{C_\tau<0} ((\varepsilon_{I,\tau} - C_\tau)^- + \rec'_C (\varepsilon_{I,\tau} - C_\tau)^+) \medskip\\
& \quad + \ind{\tau=\tau_C<T} D(t,\tau) \ind{\varepsilon_{I,\tau}>0} \ind{C_\tau>0} ((\varepsilon_{I,\tau} - C_\tau)^- + \rec_C(\varepsilon_{I,\tau} - C_\tau)^+) \medskip\\
& \quad + \ind{\tau=\tau_C<T} D(t,\tau) \ind{\varepsilon_{I,\tau}>0} \ind{C_\tau<0} (\rec_C \varepsilon_{I,\tau} - \rec'_C C_\tau) \medskip\\
& \quad + \ind{\tau=\tau_I<T} D(t,\tau) \ind{\varepsilon_{C,\tau}>0} \ind{C_\tau<0} (\varepsilon_{C,\tau} - C_\tau) \medskip\\
& \quad + \ind{\tau=\tau_I<T} D(t,\tau) \ind{\varepsilon_{C,\tau}>0} \ind{C_\tau>0} ((\varepsilon_{C,\tau} - C_\tau)^+ + \rec'_I (\varepsilon_{C,\tau} - C_\tau)^-) \medskip\\
& \quad + \ind{\tau=\tau_I<T} D(t,\tau) \ind{\varepsilon_{C,\tau}<0} \ind{C_\tau<0} ((\varepsilon_{C,\tau} - C_\tau)^+ + \rec_I(\varepsilon_{C,\tau} - C_\tau)^-) \medskip\\
& \quad + \ind{\tau=\tau_I<T} D(t,\tau) \ind{\varepsilon_{C,\tau}<0} \ind{C_\tau>0} (\rec_I \varepsilon_{C,\tau} - \rec'_I C_\tau)
\end{split}
\]
%}%
By a straightforward calculation we get
%{\small
\[
\begin{split}
{\Pi^D}(t,T;C) & = \Pi(t,T) \medskip\\
 & \quad - \ind{\tau<T} D(t,\tau) \left( \Pi(\tau,T) - \ind{\tau=\tau_C} \varepsilon_{I,\tau} - \ind{\tau=\tau_I} \varepsilon_{C,\tau} \right) \medskip\\
 & \quad - \ind{\tau=\tau_C<T} D(t,\tau) (1-\rec_C) (\varepsilon_{I,\tau}^+ - C_\tau^+)^+ \medskip\\
 & \quad - \ind{\tau=\tau_C<T} D(t,\tau) (1-\rec'_C) (\varepsilon_{I,\tau}^- - C_\tau^-)^+ \medskip\\
 & \quad - \ind{\tau=\tau_I<T} D(t,\tau) (1-\rec_I) (\varepsilon_{C,\tau}^- - C_\tau^-)^- \medskip\\
 & \quad - \ind{\tau=\tau_I<T} D(t,\tau) (1-\rec'_I) (\varepsilon_{C,\tau}^+ - C_\tau^+)^-
\end{split}
\]
%}%
Notice that the collateral account enters only as a term reducing the exposure of each party upon default of the other one, taking into account which party posted the collateral.

\subsubsection*{BCCVA General Formula}

As last step we introduce the {\emph{mid-market mark-to-market exposure}} $\varepsilon_u$ with $t \le u \le T$ as given by
\[
\varepsilon_u := \Ex{u}{\Pi(u,T)}
\;,\quad
t \le u \le T
\]
which represents the risk-free price of all cash flows remaining after time $u$ up to maturity $T$. Hence, by taking risk-neutral expectation of both sides of the equation expressing BCVA, and by plugging in the definition of mid-market exposure, we obtain the general expression for collateralized bilateral CVA.
%{\small
\begin{equation}
\begin{split}
 \brccva(t,T;C) & = - \Ex{t}{\ind{\tau<T} D(t,\tau) \left( \varepsilon_\tau - \ind{\tau=\tau_C} \varepsilon_{I,\tau} - \ind{\tau=\tau_I} \varepsilon_{C,\tau} \right)} \medskip\\
 & \quad - \Ex{t}{\ind{\tau=\tau_C<T} D(t,\tau) \lgd_C (\varepsilon_{I,\tau}^+ - C_\tau^+)^+} \medskip\\
 & \quad - \Ex{t}{\ind{\tau=\tau_C<T} D(t,\tau) \lgd'_C (\varepsilon_{I,\tau}^- - C_\tau^-)^+} \medskip\\
 & \quad - \Ex{t}{\ind{\tau=\tau_I<T} D(t,\tau) \lgd_I (\varepsilon_{C,\tau}^- - C_\tau^-)^-} \medskip\\
 & \quad - \Ex{t}{\ind{\tau=\tau_I<T} D(t,\tau) \lgd'_I (\varepsilon_{C,\tau}^+ - C_\tau^+)^-}
\end{split}
\label{eq:generalformCVA}
\end{equation}
%}

The first term on right-hand side of the above equation represents the mismatch in calculating mid-market mark-to-market exposure and on-default exposures. The second and third terms are the counterparty risk due to counterparty's default (also known as counterparty valuation adjustment or CVA), and come with a negative sign (always from the point of view of the investor). The fourth and fifth terms represent the counterparty risk due to investor's default (also known as debit valuation adjustment or DVA) and come with a positive sign (again from the point of view of the investor).

\subsubsection*{CCVA and CDVA Definitions}

We may introduce Collateralized Credit Valuation Adjustment (CCVA) and Collateralized Debit Valuation Adjustment (CDVA), and rewrite the general expression for collateralized bilateral CVA as
\[
\begin{split}
 \brccva(t,T;C)
 = & - \Ex{t}{\ind{\tau<T} D(t,\tau) \left( \varepsilon_\tau - \ind{\tau=\tau_C} \varepsilon_{I,\tau} - \ind{\tau=\tau_I} \varepsilon_{C,\tau} \right)} \medskip\\
   & - \ccva(t,T;C) \medskip\\
   & + \cdva(t,T;C)
\end{split}
\]%
where
\[
\begin{split}
 \ccva(t,T;C)
 := & \; \Ex{t}{\ind{\tau=\tau_C<T} D(t,\tau) \lgd_C (\varepsilon_{I,\tau}^+ - C_\tau^+)^+} \medskip\\
    & \; \Ex{t}{\ind{\tau=\tau_C<T} D(t,\tau) \lgd'_C (\varepsilon_{I,\tau}^- - C_\tau^-)^+}
\end{split}
\]%
and
\[
\begin{split}
 \cdva(t,T;C)
 := & - \Ex{t}{\ind{\tau=\tau_I<T} D(t,\tau) \lgd_I (\varepsilon_{C,\tau}^- - C_\tau^-)^-} \medskip\\
    & - \Ex{t}{\ind{\tau=\tau_I<T} D(t,\tau) \lgd'_I (\varepsilon_{C,\tau}^+ - C_\tau^+)^-}
\end{split}
\]

\begin{remark} {\bf(CCVA/CDVA vs. Collateral Adjusted UCVA)}
Notice that CCVA is not the unilateral CVA adjusted for collateral as seen from the investor ``I" when assuming only the counterparty ``C" may default, since this is also driven by the default time of the investor itself. Similarly, CDVA is not simply the unilateral CVA adjustment seen from the point of view of the counterparty when assuming that only the investor may default, since it contains also the counterparty default time.
\end{remark}

\subsection{Close-Out Amount Evaluation}
\label{subs:closeoutEval}

The ISDA Market Review of OTC Derivative Bilateral Collateralization Practices (2010) on section 2.1.5. states the following:
\begin{quotation}{\it
Upon default close-out, valuations will in many circumstances reflect the replacement cost of transactions calculated at the terminating party's bid or offer side of the market, and will often take into account the credit-worthiness of the terminating party. However, it should be noted that exposure is calculated at mid-market levels so as not to penalize one party or the other. As a result of this, the amount of collateral held to secure exposure may be more or less than the termination payment determined upon a close-out.
}\end{quotation}

The close-out amount is defined by the ISDA Master Agreement either as a replacement cost or as an economic equivalent of the terminated transaction, by acting in ``good faith'' and by using ``commercially reasonable'' procedures. Notice that the choice on how to compute the on-default exposure is left to the surviving party and there is not a clear statement on evaluation timing schedule. Indeed, the surviving party may require several days to complete the close-out procedure. We refer to \cite{ParkerMcGarry} for a detailed description of failings and issues related to the close-out amount evaluation procedure.

\begin{remark}{\bf (Margin Period of Risk)}
The time elapsed between the default event and the completion of the close-out procedure is named the \emph{margin period of risk}. The \cite{basel3} warns to increase the margin period of risk to capture the illiquidity of collateral and trades, the length of margin call disputes, as well as the costs of trade replacement and operations, in order to avoid exposure underestimates. For instance, they say that, if the trade involves illiquid collateral, or derivative that cannot be easily replaced, the margin period of risk should be equal to the collateral margining update interval plus 20 days.
\end{remark}

The ISDA Market Review continues with
\begin{quotation}{\it
Other differences in the valuation methodologies applied to the determination of any payment on early termination also contribute to the potential for discrepancy between these two amounts. A party may take into account the costs of terminating, liquidating or re-establishing any hedge or related
trading position. Further, it will also be reasonable to consider the cost of funding.
}\end{quotation}

The on-default exposure depends on many other factors besides the credit-worthiness of the surviving party. If we start considering such effects, we should add also the funding costs for our trading and collateral positions as well. In particular while determining a close-out amount, the determining party may consider any relevant information, including:
\begin{enumerate}
\item quotations (either firm or indicative) for replacement transactions supplied by one or more third parties that may take into account the
credit-worthiness of the determining party;
\item informations consisting of relevant market data; or
\item informations from internal sources if used by the determining party in the regular course of its business for the valuation of similar transactions.
\end{enumerate}

Such broad framework prevents to achieve a tight definition of close-out amount or of on-default exposure, and it can clearly produce a wide range of results. See, for instance, \cite{WeeberRobson} where the authors show realistic examples on how evaluating the close-out amount.

Yet, if we disregard the issues coming from loosely defined terms, we could approximate on-default exposures $\varepsilon_{I,\tau_C}$ and $\varepsilon_{C,\tau_I}$ with the value a of replacement operation with a risk-free counterparty (with the same collateralization rule), as shown in \cite{pallaNested}. Hence, if we apply our collateralized bilateral CVA formula to the risk-free payoff to include the credit-worthiness of the surviving party, we get
\[
\begin{split}
\varepsilon_{I,\tau_C} \doteq \varepsilon_{\tau_C}
 & - \Ex{t}{\ind{\tau_C<\tau_I<T} D(\tau_C,\tau_I) \lgd_I (\varepsilon_{\tau_I}^- - C_{\tau_I}^-)^-} \medskip\\
 & - \Ex{t}{\ind{\tau_C<\tau_I<T} D(\tau_C,\tau_I) \lgd'_I (\varepsilon_{\tau_I}^+ - C_{\tau_I}^+)^-}
\end{split}
\]%
and
\[
\begin{split}
\varepsilon_{C,\tau_I} \doteq \varepsilon_{\tau_I}
 & - \Ex{t}{\ind{\tau_I<\tau_C<T} D(\tau_I,\tau_C) \lgd_C (\varepsilon_{\tau_C}^+ - C_{\tau_C}^+)^+} \medskip\\
 & - \Ex{t}{\ind{\tau_I<\tau_C<T} D(\tau_I,\tau_C) \lgd'_C (\varepsilon_{\tau_C}^- - C_{\tau_C}^-)^+}
\end{split}
\]

The above treatment of close-out amounts is also known as {\emph {nested bilateral CVA}} due to the nested application of CVA and DVA formulas. A detailed discussion on the effects of employing such approximation can be found in \cite{BrigoMorini}.

\subsection{Special Cases of Bilateral Collateralized CVA}

In this section, we specialize the general CVA formula given in Eq.~(\ref{eq:generalformCVA}). We start showing the formula in the case when all the exposures are evaluated at mid-market, i.e. we consider:
\[
\varepsilon_{I,t} = \varepsilon_{C,t} = \varepsilon_t
\]
We then obtain that the collateralized bilateral CVA is equal to
\begin{equation}
\begin{split}
  \brccva(t,T;C) = & - \Ex{t}{\ind{\tau=\tau_C<T} D(t,\tau) \lgd_C (\varepsilon_\tau^+ - C_\tau^+)^+} \medskip\\
                   & - \Ex{t}{\ind{\tau=\tau_C<T} D(t,\tau) \lgd'_C (\varepsilon_\tau^- - C_\tau^-)^+} \medskip\\
                   & - \Ex{t}{\ind{\tau=\tau_I<T} D(t,\tau) \lgd_I (\varepsilon_\tau^- - C_\tau^-)^-} \medskip\\
                   & - \Ex{t}{\ind{\tau=\tau_I<T} D(t,\tau) \lgd'_I (\varepsilon_\tau^+ - C_\tau^+)^-}
\end{split}
\end{equation}

If collateral re-hypothecation is not allowed ($\lgd'_C = \lgd'_I = 0$), then the above formula simplifies to
\begin{equation}
\label{eq:CVAnorehp}
\begin{split}
\brccva(t,T;C) = & - \Ex{t}{\ind{\tau=\tau_C<T} D(t,\tau) \lgd_C (\varepsilon_\tau^+ - C_\tau^+)^+} \medskip\\
                 & - \Ex{t}{\ind{\tau=\tau_I<T} D(t,\tau) \lgd_I (\varepsilon_\tau^- - C_\tau^-)^-}
\end{split}
\end{equation}

On the other hand, if re-hypothecation is allowed and the surviving party always faces the worst case ($\lgd'_C = \lgd_C$ and $\lgd'_I = \lgd_I$), then we get

\begin{equation}
\label{eq:CVArehp}
\begin{split}
\brccva(t,T;C) = & - \Ex{t}{\ind{\tau=\tau_C<T} D(t,\tau) \lgd_C (\varepsilon_\tau - C_\tau)^+} \medskip\\
                 & - \Ex{t}{\ind{\tau=\tau_I<T} D(t,\tau) \lgd_I (\varepsilon_\tau - C_\tau)^-}
\end{split}
\end{equation}

Finally, if we remove collateralization, i.e. $C_t=0$, then we recover the result of \cite{BrigoCap10} used in \cite{BrigoPallPap}
\begin{equation}
\begin{split}
\brcva(t,T) = & - \Ex{t}{\ind{\tau=\tau_C<T} D(t,\tau) \lgd_C \varepsilon_\tau^+} \medskip\\
              & - \Ex{t}{\ind{\tau=\tau_I<T} D(t,\tau) \lgd_I \varepsilon_\tau^-}
\end{split}
\end{equation}

If we remove collateralization ($C_t=0$) and we consider a risk-free investor ($\tau_I\rightarrow\infty$), we recover the result of
\cite{BrigoPall07}, see also \cite{CannDuff}.
\begin{equation}
 \cva(t,T) = - \Ex{t}{\ind{\tau_C<T} D(t,\tau_C) \lgd_C \varepsilon_{\tau_C}^+}
\end{equation}

\section{Example of Collateralization Schemes}
\label{sec:examples}

We consider a setting where investor and counterparty exposures equal the mid-market mark-to-market exposure, where there are no funding costs for either party, and where re-hypothecation is not allowed. Therefore, the resulting CVA formula is given by Eq.~(\ref{eq:CVAnorehp}). We consider two collateralization mechanisms.

The first mechanism  removes all the exposure risk of the parties and is therefore called {\emph {perfect collateralization}}.

The second mechanism, instead, is the most realistic and follows the margining practice where during the life of the deal both parties post or withdraw collateral on a fixed set of dates, according to their current exposure, to or from an account held by the Collateral Taker.
In general, the collateral taker may be a third party or the party of the transaction who is not posting collateral. We call the second mechanism {\emph {collateralization through margining}}.

\subsection{Perfect Collateralization}

The perfect collateralization scheme consists in updating the collateral account continuously, thereby obtaining the following collateralization rule.
\[
C^{\rm perfect}_t := \varepsilon_t
\]

Thus, if we plug it into the collateralized bilateral CVA equation, we get that all terms drop, as expected, leading to
\[
\brccva(t,T;C^{\rm perfect}) = 0
\]
and
\[
\Ex{t}{{\Pi^D}(t,T;C)} = \Ex{t}{\Pi(t,T)} = \varepsilon_t = C^{\rm perfect}_t
\]

Under this collateralization rule, the proper discount curve for pricing the deal is the collateral accrual curve, see \cite{Fujii2010} and \cite{Piter2010}.

\subsection{Collateralization through Margining}

We assume that collateral posting only occurs at discrete times on a fixed grid ($t_0=t,\ldots,t_N=T$), and we allow for the presence of minimum transfer amounts ($M > 0$), and thresholds ($H$), with $H\ge M$. Thresholds represent the amount of permitted unsecured risk, so that they may depend on the credit quality\footnote{Moving thresholds depending on a deterioration of the credit quality of the counterparties (downgrade triggers) have been a source of liquidity strain during the market crisis. See BIS white paper The role of margin requirements and haircuts in procyclicality (2010).} of the counterparties.

A realistic margining practice also includes independent amounts, which represent a further insurance on the transaction and they are often posted as an upfront protection, but they may be updated according to exposure changes. We do not consider independent amounts in the following.

At each collateral posting date $t_i$, the collateral account is updated according to changes in exposure. We denote by $C_{t_i^-}$ the collateral account right before the collateral update for time $t_i$ takes place, and denote by $C_{t_i^+}$ the collateral account right after the collateral update for time $t_i$ takes place. We first consider how much collateral the investor
should post to or withdraw from the collateral account. This is given by

\begin{equation}\label{eq:collpostI}
\ind{|(\varepsilon_{t_i} + H_I)^- - C^-_{t_i^-}| > M} ((\varepsilon_{t_i} + H_I)^- - C^-_{t_i^-})
\end{equation}
Then, we consider how much collateral the counterparty should post to or withdraw from the collateral account. This is given by
\begin{equation}\label{eq:collpostC}
\ind{|(\varepsilon_{t_i} - H_C)^+ - C^+_{t_i^-}| > M} ((\varepsilon_{t_i} - H_C)^+ - C^+_{t_i^-})
\end{equation}

We have
\[
C_{t_0} := 0 \;,\quad C_{t_n^+} := 0 \;,\quad C_{u^-} := \frac{C_{{\beta(u)}^+}}{D(\beta(u),u)}
\]
and
\begin{equation}\label{eq:collupd}
\begin{split}
C_{t_i^+} := C_{t_i^-} & + \ind{|(\varepsilon_{t_i} + H_I)^- - C^-_{t_i^-}| > M} ((\varepsilon_{t_i} + H_I)^- - C^-_{t_i^-}) \medskip\\
                       & + \ind{|(\varepsilon_{t_i} - H_C)^+ - C^+_{t_i^-}| > M} ((\varepsilon_{t_i} - H_C)^+ - C^+_{t_i^-})
\end{split}
\end{equation}
where $\beta(u)$ is the last update time before $u$, and $t_0 < u \le t_n$.

We are also implicitly assuming that, on default occurrence at time $t_i$, all collateral requests initiated, but not yet completed, are neglected.

In case of no thresholds ($H_I=H_C=0$) and no minimum transfer amount ($M=0$), we obtain a simpler rule
\[
C_{t_0} = C_{t_n^+} = 0
\;,\quad
C_{t^-} = \frac{\varepsilon_{\beta(u)}}{D(\beta(u),u)}
\;,\quad
C_{t_i^+} = \varepsilon_{t_i}
\]

\section{Application to Interest-Rate Swaps}
\label{sec:IRS}

In this section we extend the analysis of \cite{BrigoPallPap} by presenting some numerics on the collateralized CVA.

First we briefly summarize model assumptions: we consider a model that is stochastic both in the interest rates (underlying market) and in the default intensities (investor and counterparty defaults). Joint stochasticity is needed to introduce correlation between rates and credit. The interest-rate sector is modeled according to a short-rate Gaussian shifted two-factor process (hereafter G2++), while each of the two default-intensity sectors is modeled according to a shifted square-root diffusion process (hereafter SSRD). Details on the G2++ model can be found in \cite{BrigoPallPap}, whereas for the SSRD model we refer to \cite{BrigoAlf05} and \cite{BrigoElBachir}. The two models are coupled by correlating their Brownian shocks.

\subsection{Interest-Rate Model}
\label{subsec:interestrates}

For interest rates, we assume that the dynamics of the instantaneous short-rate process under the risk-neutral measure is given by
\begin{equation}
\label{ch4:hw2srdyn}
r(t) = x(t) + z(t) + \varphi(t;\alpha) \;,\quad r(0)=r_0,
\end{equation}%
where $\alpha$ is a set of parameters and the processes $x$ and $z$ are ${\cal F}_t$ adapted and satisfy
\begin{equation}
\label{ch4:hw2xydyn}
\begin{split}
dx(t) & = -a x(t) \,dt + \sigma \,dZ_1(t) \;,\quad x(0)=0,\\
dz(t) & = -b z(t) \,dt + \eta \,dZ_2(t) \;,\quad z(0)=0,
\end{split}
\end{equation}%
where $(Z_1,Z_2)$ is a two-dimensional Brownian motion with instantaneous correlation $\rho_{12}$ as from
\[
d \langle Z_1, Z_2\rangle_t = \rho_{12} \,dt,
\;,\quad
-1 \le \rho_{12} \le 1
\]%
and $r_0$, $a$, $b$, $\sigma$, $\eta$ are positive constants. The model can be extended to include time-dependend volatilities as shown in \cite{BrigoPallPap}.

We calibrate the interest-rate model parameters to the initial zero coupon curve observed in the market and to the at-the-money swaption volatilities quoted by the market on May 26, 2009. Market data and calibrated model parameters can be found in \cite{BrigoPallPap}, while more details on the methodology can be found in \cite{BrigoPall07}.

\subsection{Counterparty and Investor Credit-Spread Models}
\label{subsec:creditspreads}

For the stochastic intensity models we set
\begin{equation}
\label{extended_i}
\lambda^{i}_t = y^{i}_t + \psi^{i}(t;\beta^i) \;,\quad i\in \{I,C\}
\end{equation}
where whenever we omit the upper index we refer to quantities for both indices. Here the index value``I" denotes the investor and ``C" the counterparty. The function $\psi$ is a deterministic function, depending on the parameter vector $\beta$ (which includes $y_0$), that is integrable on closed intervals. The initial condition $y_0$ is one more parameter at our disposal. We are free to select its value as long as%
\[
\psi^i(0;\beta) = \lambda^i_0 - y^i_0  \;,\quad i \in \{I,C\} \, .
\]
We take $y$ to be a Cox Ingersoll Ross process as given by
\[
dy^i_t = \kappa^i(\mu^i-y^i_t)\,dt + \nu^i \sqrt{y^i_t}\,dZ^i_3(t) \;,\quad i \in \{I,C\}
\]
where the parameter vector is $\beta^i \eqdef (\kappa^i,\mu^i,\nu^i,y^i_0)$ and each parameter is a positive deterministic constant. As usual, $Z_3^i$ is a standard Brownian motion process under the risk neutral measure. The two processes $y^I$ and $y^C$ are assumed to be independent, so that $Z_3^I$ is independent of $Z_3^C$. Correlation between defaults of $I$ and $C$ will be introduced as a pure default correlation below. This independence for the spreads is assumed to simplify the parametrization of the model and focus on default correlation rather than spread correlation, but the assumption can be removed if one is willing to complicate the parametrization of the model. The model can be extended to incorporate jumps as explained in \cite{BrigoPall07}.

We calibrate our model to two different sets of CDS spreads and implied volatilities, which we name hereafter {\em Mid} and {\em High} risk settings. The former set consists of smaller CDS spreads and volatilies than the latter one. More details on the calibration procedure can be found in \cite{BrigoPallPap}.

\subsection{Correlation Parameters}

First, we take the short interest-rate factors $x$ and $z$ and the intensity process $y$ to be correlated, by assuming the driving Brownian motions $Z_1, Z_2$ and $Z_3$ to be instantaneously correlated according to
\[
d \langle Z_j ,Z^i_3 \rangle_t  = \rho_{j,i} \,dt \;,\quad j \in \{1,2\} \;,\quad i \in \{I,C\}
\]
The instantaneous correlation between the resulting short-rate and the intensity, i.e. the instantaneous interest-rate / credit-spread correlation is constant and given by
\[
\bar{\rho}_i = \frac{\sigma \rho_{1i}+\eta \rho_{2i}}{\sqrt{\sigma^2 + \eta^2 + 2 \sigma \eta \rho_{12}}} \ .
\]

Then, concerning default events, we prefer to model default correlation by introducing a Gaussian copula on default times, rather than by correlating the default intensities. Thus, we define the cumulative intensities
\[
\Lambda^i(0,t) := \int_0^t\lambda^i(u)\,du \;,\quad i \in \{I,C\}
\]
and we simulate the survival indicator of each name by sampling them as
\[
\ind{\tau^k<t} = \ind{U^k>\exp\{\Lambda^k(0,t)\}} \;,\quad U^k := \Phi(z^k) \;,\quad k \in \{I,C\}, \ \ (z^C,z^I) \sim {\cal N}_2(\rho_G)
\]%
where $\Phi$ is the standard normal cumulative distribution function, and ${\cal N}_2(\rho_G)$ is the bivariate standard normal distribution with correlation parameter $\rho_G$.

Furthermore, in order to reduce the number of free parameters and to model the correlation structure in a more robust way, in the following we assume that
\[
\rho_{1i} = \rho_{2i} =: \bar{\rho}_i
\;,\quad i \in \{I,C\} \ .
\]

Hence, we have as free correlation parameters only ${\bar\rho}_C$, ${\bar\rho}_I$ and $\rho_G$, recovering the other correlations from them.

\subsection{Numerical Examples}
\label{sec:casestudyrates}

We use the collateralization mechanism through margining described in Section \ref{sec:examples}. We assume zero minimum transfer amount and thresholds $M=H_I=H_C=0$. Under this collateralization mechanism, we consider the behavior of the bilateral credit valuation adjustment as a function of $\delta$, where $\delta := t_i - t_{i-1}$ is the time between two consecutive collateral update times. We consider both the case when received collateral cannot be re-hypothecated by the collateral taker (BCCVA given by Eq.~\ref{eq:CVAnorehp}) and and the case when it can be re-hypothecated and the surviving party always faces the worst case (BCCVA given by Eq.~\ref{eq:CVArehp}).

\subsubsection*{Changing the Margining Frequency}
				
First, we consider the margining frequency $\delta$ ranging from one week up to six months. Notice that we are considering interest-rate swaps (IRS) with one-year payment frequency for the fix leg and six-month frequency for the floating leg (as usually found in the Euro market ). By keeping the frequency $\delta$ below six month, we avoid jumps in BCCVA occurring at the times when cash-flows are exchanged.

\begin{figure}
\begin{center}
\includegraphics[scale=0.75]{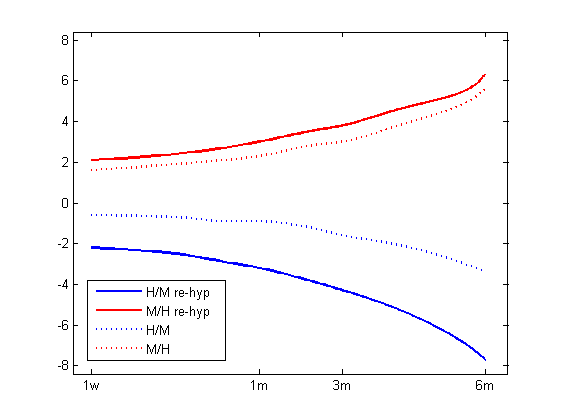} \\
\includegraphics[scale=0.45]{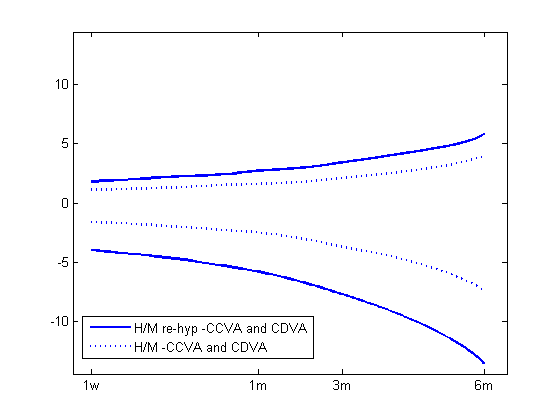}
\includegraphics[scale=0.45]{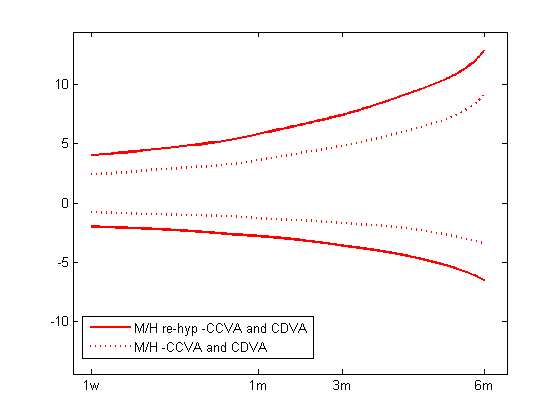}
\caption{BCCVA for a ten-year IRS under collateralization through margining as a function of the update frequency $\delta$ with ${\bar\rho}_C={\bar\rho}_I=\rho_G=0$. Update frequencies under six months. Continuous lines represent the re-hypothecation case, while dotted lines represent the opposite case. The red line represents an investor riskier than the counterparty (mid-risk counterparty and high-risk investor, or M/H), while the blue line represents an investor less risky than the counterparty (high-risk counterparty and mid-risk investor, or H/M). The upper panel plots the BCCVA, while the bottom left and right panels plot respectively the $-$CCVA and CDVA components. All values are in basis points.}
\label{fig:cvaUpdate6m}
\end{center}
\end{figure}

In figure \ref{fig:cvaUpdate6m} we show the sensitivity of the BCCVA of an IRS with ten years maturity to the update frequency of collateral margining, which ranges from one week to six months. We see that the case of an investor riskier than the counterparty leads to positive value for BCCVA, while the case of an investor less risky than the counterparty has the opposite behaviour. In order to better explain that, we also plot separately the $-$CCVA and CDVA terms contributing to the adjustment. It is evident from the figure that, when the investor is riskier the CDVA part of the correction dominates, while when the investor is less risky the counterparty has the opposite behaviour. The effect of re-hypothecation is to enhance the absolute size of the correction, a reasonable behaviour, since, in such case, each party has a greater risk because of being unsecured on the collateral amount posted to the other party in case of default.

Although realistic update frequencies are usually weekly or daily, and only in exceptional cases reach the order of some months,  we also plot all the cases from 1 to ten years (namely no margining at all) as a tool to discuss collateral re-hypothecation effects.

\begin{figure}
\begin{center}
\includegraphics[scale=0.75]{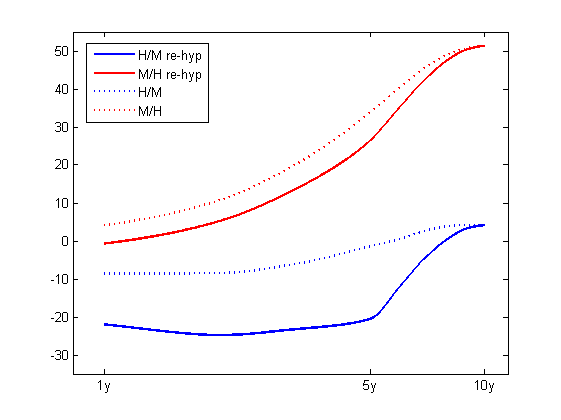}\\
\includegraphics[scale=0.45]{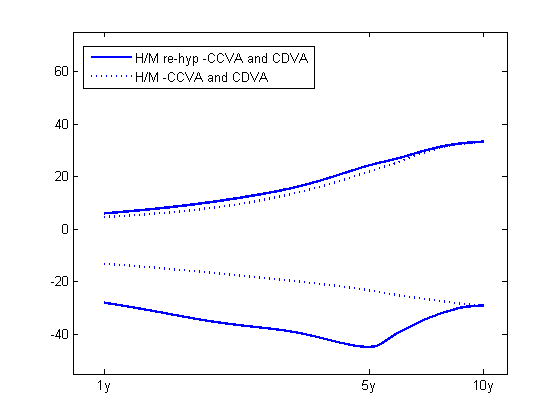}
\includegraphics[scale=0.45]{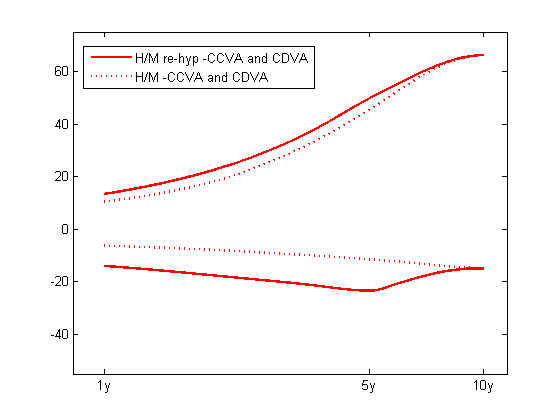}
\caption{BCCVA for a ten-year IRS under collateralization through margining as a function of the update frequency $\delta$ with ${\bar\rho}_C={\bar\rho}_I=\rho_G=0$. Update frequencies from one to ten years. Continuous lines represent the re-hypothecation case, while dotted lines represent the opposite case. The red line represents an investor riskier than the counterparty (mid-risk counterparty and high-risk investor, or M/H), while the blue line represents an investor less risky than the counterparty (high-risk counterparty and mid-risk investor, or H/M). The upper panel plots the BCCVA, while the bottom left and right panels plot respectively the $-$CCVA and CDVA components. All values in basis points.}
\label{fig:cvaUpdate1y}
\end{center}
\end{figure}

\subsubsection*{Inspecting the Exposure Profiles}

\begin{figure}
\begin{center}
\includegraphics[scale=0.3]{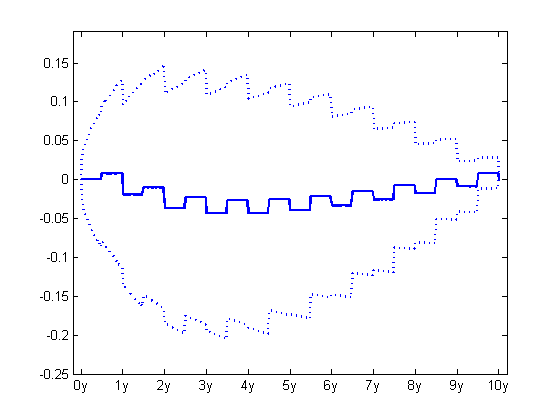}
\includegraphics[scale=0.3]{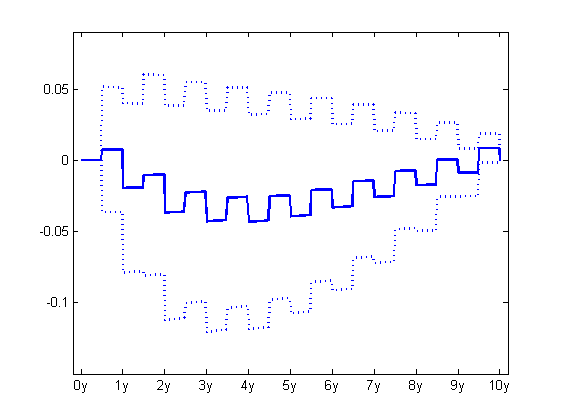}
\includegraphics[scale=0.3]{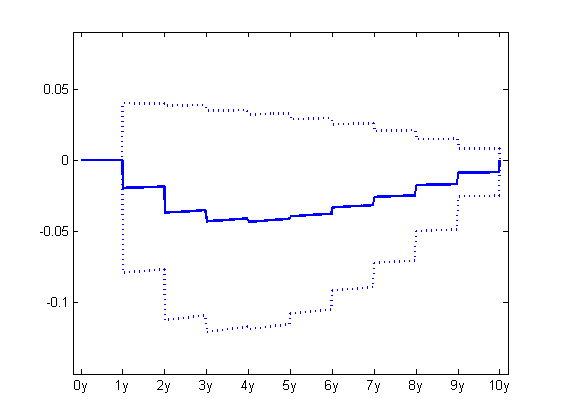}
\caption{Left panel: mark-to-market density of a ten-year IRS uncollateralized exposure through time. Mid panel: collateral density for six-month update frequency through time. Right panel: collateral density for one-year update frequency through time. Continuous lines are distributions' mean values, while dotted lines are $95$ percentiles.}
\label{fig:cvaMTM}
\end{center}
\end{figure}

If we look at figure \ref{fig:cvaUpdate1y}, namely update frequencies greater than one year, we observe that the case of an investor riskier than the counterparty has a greater BCCVA without re-hypothecation. The opposite occurs for frequency under six months. The explanation is due to the fact that the preceding reasoning holds separately for CCVA and CDVA, and not for their difference. Indeed, when the update frequency is equal to one year or greater the investor has a greater probability of posting collateral, as shown in figure \ref{fig:cvaMTM}, leading to an increase in CCVA when re-hypothecation is allowed, but CDVA is little affected.

Further insights can be gained by looking at the expected exposure profiles which contribute to the BCCVA adjustment. Here, we differentiate between
\begin{itemize}
\item the positive part of the (uncollateralized) exposure $\varepsilon_u^+$ and its negative part $\varepsilon_u^-$;
\item the collateralized expected exposure without re-hypothecation contributing to the CCVA adjustment
$(\varepsilon_u^+ - C_u^+)^+$ and the corresponding term $(\varepsilon_u^- - C_u^-)^-$ contributing to the CDVA adjustment.
\item the collateralized expected exposure with re-hypothecation contributing to the CCVA adjustment $(\varepsilon_u - C_u)^+$ and the corresponding term $(\varepsilon_u - C_u)^-$ contributing to the CDVA adjustment.
\end{itemize}

\begin{figure}
\begin{center}
\includegraphics[scale=0.45]{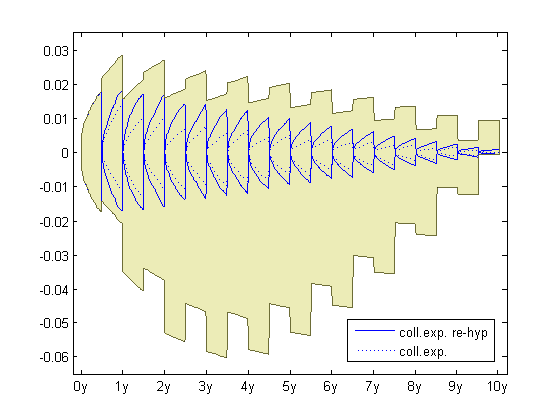}
\includegraphics[scale=0.45]{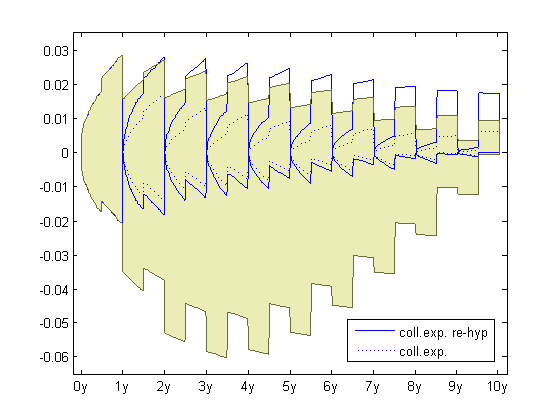}
\caption{Expected exposure profiles for a ten-year IRS through time. The borders of the yellow area are the mean values of the positive and negative parts of the uncollateralized exposures (i.e. $\E{\varepsilon^+}$ and $\E{\varepsilon^-}$), while the blue lines are collateralized exposures (continuous line is re-hypothecation case, namely $\E{(\varepsilon_u - C_u)^+}$ and $\E{(\varepsilon_u - C_u)^-}$; dotted line the opposite case, namely $\E{(\varepsilon_u^+ - C_u^+)^+}$ and $\E{(\varepsilon_u^- - C_u^-)^-}$). Left panel: expected exposure profiles for six months collateral update frequency. Right panel: expected exposure profiles for one year collateral update frequency. All values are in basis points.}
\label{fig:exposureprof}
\end{center}
\end{figure}
We can clearly see from the right panel in figure \ref{fig:exposureprof} that, when assuming re-hypothecation, the collateralized expected exposure may exceed the uncollateralized one. This is the case because we may have $(\varepsilon_u - C_u)^+ > \varepsilon_u^+$, which holds in scenarios where at time $t$ it is more likely that $C(t) < 0$, i.e. that collateral is posted by the investor and re-hypothecated by the counterparty. Therefore, this means that the investor is now exposed to the counterparty both in terms of the mark-to-market value of the transaction, and also in terms of the posted collateral, which is an unsecured claim and may not be returned in full in case of the earlier counterparty default.

\subsubsection*{Changing the Correlation Parameters}

A second example is investigating the effects of correlations (both interest-rate/credit-spread and default-time correlations) for different frequencies for collateral update.

\begin{figure}
\begin{center}
\includegraphics[scale=0.45]{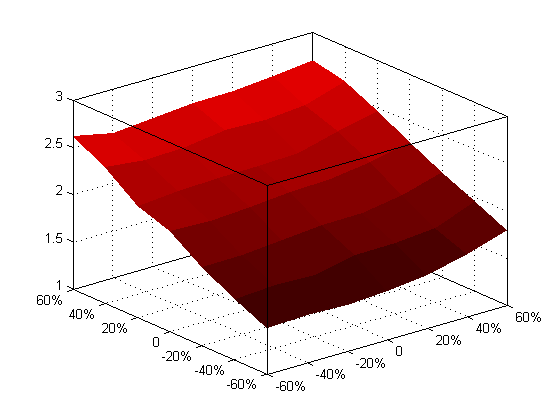}
\includegraphics[scale=0.45]{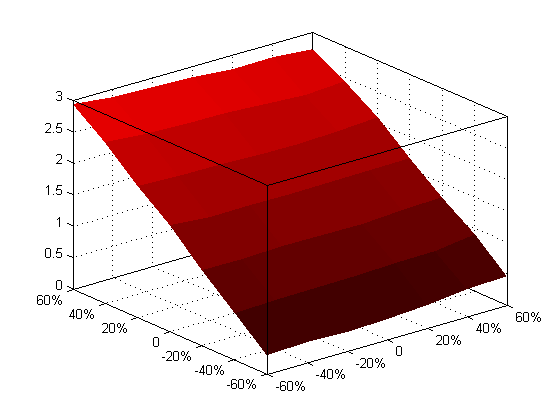} \\
\includegraphics[scale=0.45]{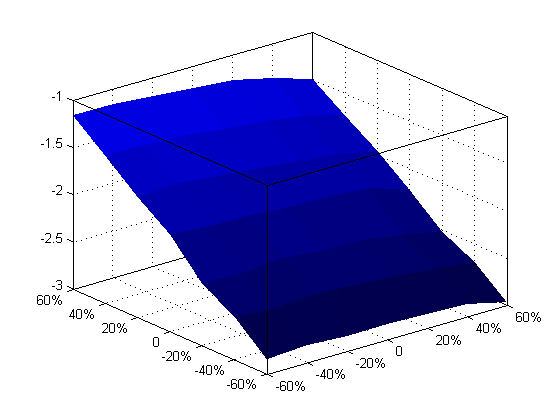}
\includegraphics[scale=0.45]{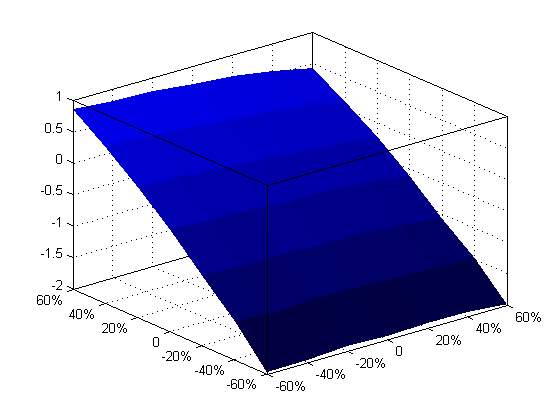}
\caption{BCCVA with collateral update frequency of one week for a ten-year IRS (M/H market settings in upper panels, H/M market settings in lower panels) with different choices of interest-rate/credit-spread correlation ($\rho_C = \rho_I$ parameters, left-side axis) and default-time correlation ($\rho_G$ Gaussian copula parameter, right-side axis). Left panels show values with re-hypothecation, while right panels without re-hypothecation. All values in basis points.
}
\label{fig:cvaRho1}
\end{center}
\end{figure}

\begin{figure}
\begin{center}
\includegraphics[scale=0.45]{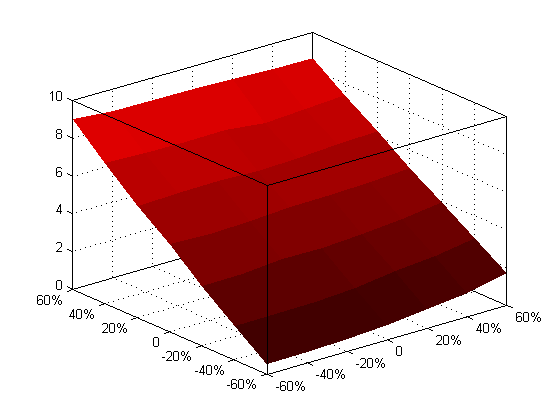}
\includegraphics[scale=0.45]{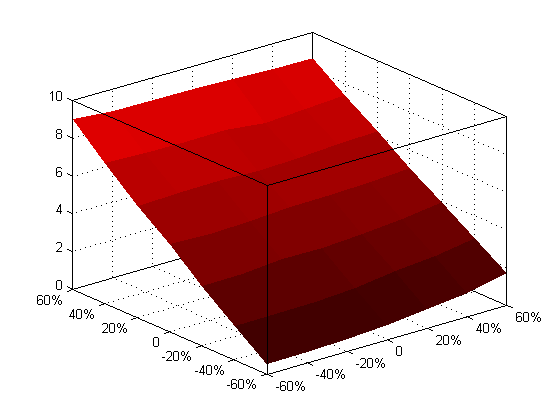} \\
\includegraphics[scale=0.45]{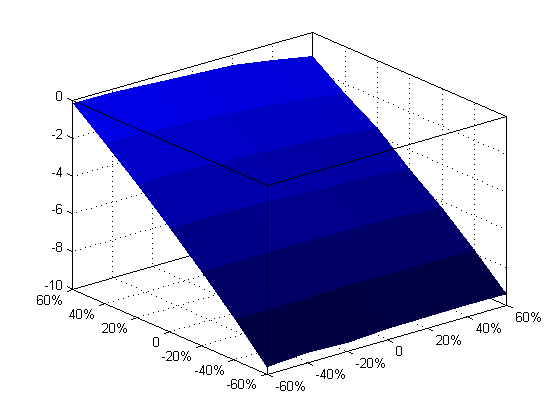}
\includegraphics[scale=0.45]{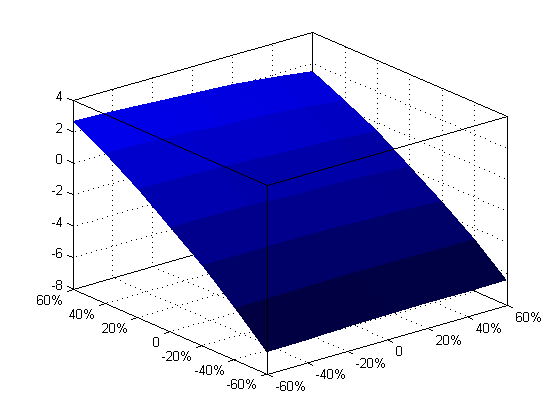}
\caption{BCCVA with collateral update frequency of three months for a ten-year IRS (M/H market settings in upper panels, H/M market settings in lower panels) with different choices of interest-rate/credit-spread correlation ($\rho_C = \rho_I$ parameters, left-side axis) and default-time correlation ($\rho_G$ Gaussian copula parameter, right-side axis). Left panels show values with re-hypothecation, while right panels without re-hypothecation. All values in basis points.
}
\label{fig:cvaRho12}
\end{center}
\end{figure}

First of all, a direct comparison between figures \ref{fig:cvaRho1} and \ref{fig:cvaRho12} shows that increasing the collateral update frequency increases the magnitude of the BCCVA adjustment (larger update frequency imply larger on-default exposures and thus larger BCCVAs), but it does not substantially change the dependence pattern of the BCCVA on the correlation parameters. Further, we notice that we get similar results both by allowing or not allowing re-hypothecation, or by chaning the market set from M/H to H/M.

We can see that, for a given level of default-time correlation parameter ${\bar\rho}_G$, a common increase in credit/interest rate correlation parameters ${\bar\rho}_C$ and ${\bar\rho}_I$ leads to higher BCCVA adjustments. This is because higher interest rates will correspond to higher credit spreads, thus putting the receiver swaption embedded in the CCVA term of the adjustment more out of the money. This will cause the CCVA term of the adjustment to diminish in absolute value, so that the final value of the bilateral CVA will be larger for high correlations. As we are considering a counterparty more risky than the investor, we will have that the CCVA term will be dominating in the adjustment over the CDVA term. This is just an example of the complexity of patterns in bilateral collateralized CVA calculations. Model dependent dynamic parameters such as volatility and correlations can change the profile of the bilateral CVA calculation even in presence of collateral

\subsubsection*{Changing the Credit-Spread Volatility}
A third example consists in changing the volatility of the credit spread and monitor the impact of wrong-way risk for different collateral update frequencies, and for different values of interest-rate/credit-spread correlation.
\begin{figure}
\begin{center}
\includegraphics[scale=0.45]{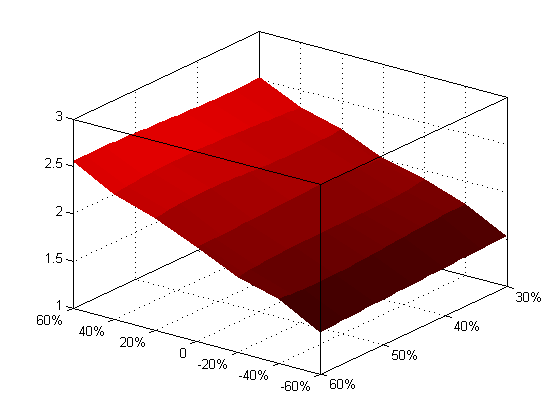}
\includegraphics[scale=0.45]{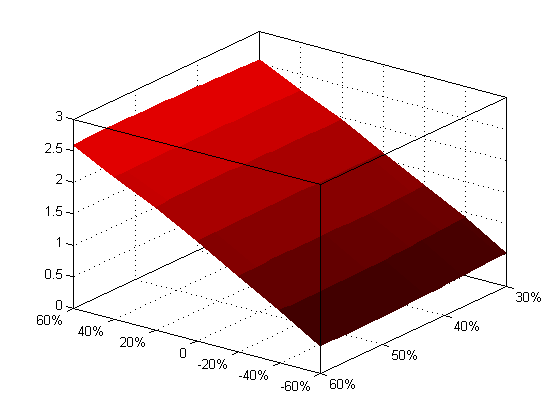} \\
\includegraphics[scale=0.45]{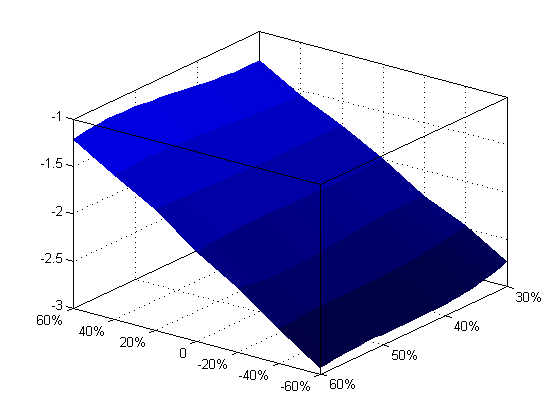}
\includegraphics[scale=0.45]{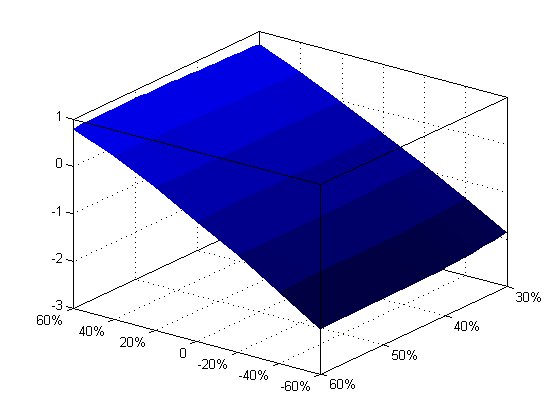}
\caption{BCCVA with collateral update frequency of one week for a ten-year IRS (M/H market settings in upper panels, H/M market settings in lower panels) with different choices of interest-rate/credit-spread correlation ($\rho_C = \rho_I$ parameters, left-side axis) and counterparty's credit-spread volatility ($\nu^C$ parameter, right-side axis). Left panels show values with re-hypothecation, while right panels without re-hypothecation. Default-time correlation $\rho_G=0$. All values in basis points.
}
\label{fig:cvaVol1}
\end{center}
\end{figure}

\begin{figure}
\begin{center}
\includegraphics[scale=0.45]{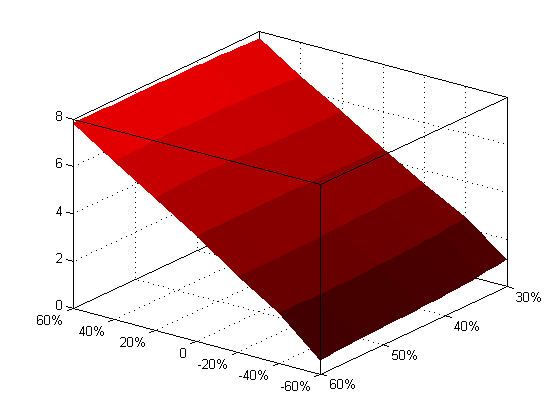}
\includegraphics[scale=0.45]{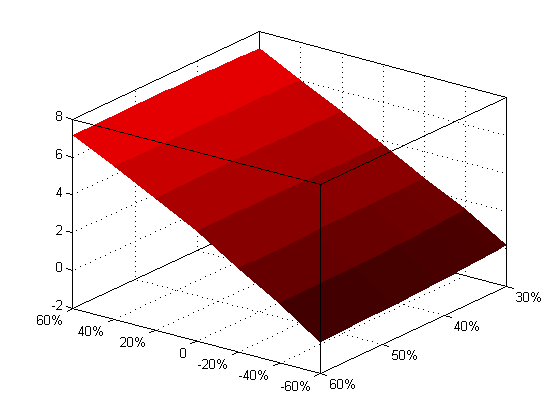} \\
\includegraphics[scale=0.45]{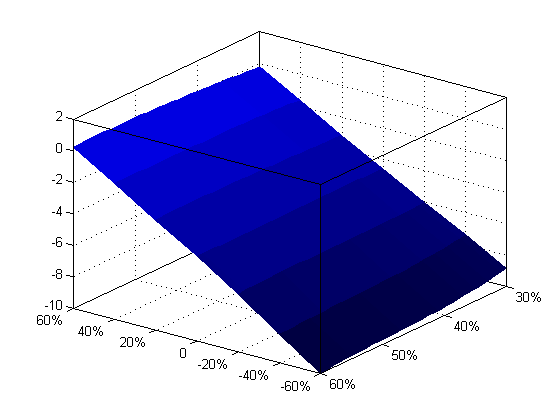}
\includegraphics[scale=0.45]{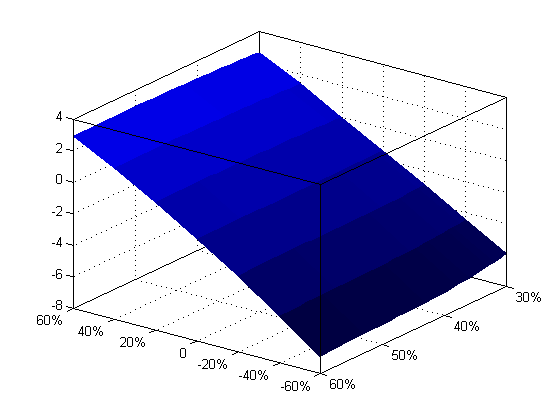}
\caption{BCCVA with collateral update frequency of three months for a ten-year IRS (M/H market settings in upper panels, H/M market settings in lower panels) with different choices of interest-rate/credit-spread correlation ($\rho_C = \rho_I$ parameters, left-side axis) and counterparty's credit-spread volatility ($\nu^C$ parameter, right-side axis). Left panels show values with re-hypothecation, while right panels without re-hypothecation. Default-time correlation $\rho_G=0$. All values in basis points.
}
\label{fig:cvaVol12}
\end{center}
\end{figure}

As in the preceding case we notice in figures \ref{fig:cvaVol1} and \ref{fig:cvaVol12} that, for a given level of the counterparty's credit-spread volatility paramter $\nu_C$, the dependency of BBCVA on the credit/interest rate correlation parameters ${\bar\rho}_C$ and ${\bar\rho}_I$ leads to higher adjustments for higher correlations. Regardless of the collateral update frequency, the credit-spread volatility has only a small impact on the BCCVA adjustment, which is much more affected by the interest-rate/credit-spread correlation. However, it is worth noticing that for different choices of ${\bar\rho}_C$, the dependence pattern of the adjustements on the credit spreads volatility may be reversed (see for instance the case when ${\bar\rho}_C = 60 \%$ where the adjustment is decreasing in $\nu_C$ and the case when ${\bar\rho}_C = -60\%$ where the adjustment is increasing in $\nu_C$).

\begin{figure}
\begin{center}
\includegraphics[scale=0.45]{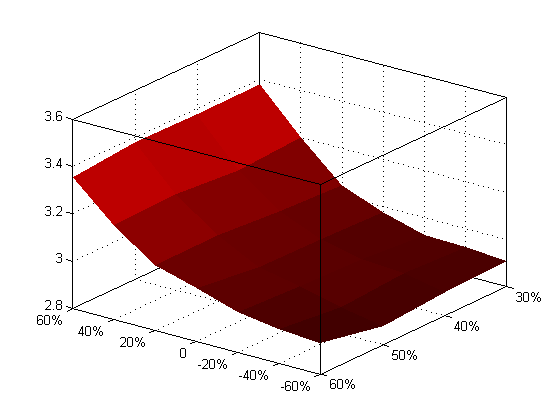}
\includegraphics[scale=0.45]{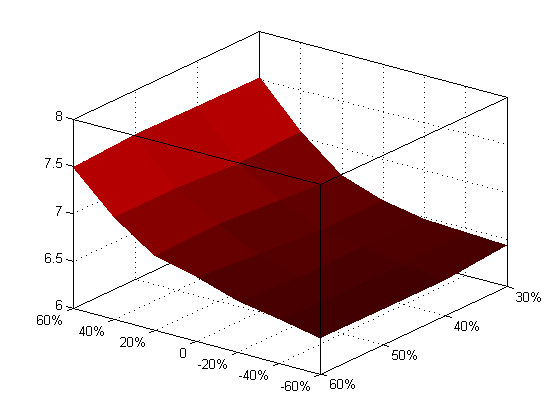} \\
\caption{BCCVA with re-hypothecation for a ten-year IRS with different choices of counterparty default correlation ($\rho_G$  parameter, left-side axis) and counterparty's credit-spread volatility ($\nu^C$ parameter, right-side axis).
We assume $\rho_C = \rho_I = 0$. The left graph refers to a collateral update frequency of one week, while the right graph refers to a collateral update frequency of three months. All values in basis points.}
\label{fig:cvaVolDef}
\end{center}
\end{figure}

We present a fourth example where we change the volatility of the credit spread and monitor the impact of wrong-way risk for different collateral update frequencies, and for different values of counterparty default correlations. We assume flat hazard rate structures, obtained as follows. We take the maximum CDS spread of the high risk name, let us call it $CDS^H$, and the maximum CDS spread in the medium risk name, let us call it $CDS^M$. We use the shift to match a flat hazard rate curve
$h_H = CDS^H/\lgd$ and $h_M = CDS^M/\lgd$. We can see from figure \ref{fig:cvaVolDef} that the adjustments tend to become higher when the default correlation is positive. This is expected because when the value of $\nu_C$ is close to zero and the correlation is positive, the party with higher hazard rate tends to default earlier than the party with smaller hazard rate in almost all default scenarios. Similarly to figure \ref{fig:cvaVol12}, notice that depending on the default correlation parameter $\rho_G$, the  dependence pattern of the credit adjustments on the credit spreads volatility may be reversed (see for instance the case when
$\rho_G = 60 \%$ where the adjustment is increasing in $\nu_C$, and compare it with the case when $\rho_G = -60\%$ where the adjustment is decreasing in $\nu_C$).

\section{Conclusions and Further Research}
\label{sec:Conclusions}

In this paper we described a complete framework for bilateral CVA risk-neutral pricing inclusive of close-out netting rules and collateral margining, considering also the case when collateral can be re-hypothecated. We consider interest-rate swap contracts and, show the impact of collateralization frequency on the bilateral CVA via numerical simulations.

%In the future, we would like to refine the definition of on-default exposures, so that it includes both credit and liquidity risk, see \cite{BrigoCapPedr10} for a survey of intensity-based frameworks modeling liquidity spreads and credit spreads into a unified framework.

In the future we would like to include even more details of the collateral margining procedure. In particular we will investigate the impact of time delay due to dispute resolutions and margin period of risk. Further, we would like to refine the definition of on-default exposures to match the ongoing work of ISDA working groups and to incorporate in our framework funding costs according to realistic liquidity policies.

Possible extesions of our framework should consider default clustering to address the systemic risk arising from the interconnectedness of banks and other financial institutions, and study the interplay between liquidity and (bilateral) counterparty risk in modelling funding curves. We would also like to address in detail the dynamic hedging of counterparty risk.

%Some efforts in this direction have already been taken in \cite{BielCrep}, although they only deal with unilateral and uncollateralized CVA.

\section*{Acknowledgments}
{
We are grateful to Marco Amicone and Andrea Germani for useful discussions on margining procedure and trading practice. We thank also Massimo Morini for discussions on nested CVA.
}

\bibliographystyle{nonumber}

\end{document}